\documentclass[12pt]{aastex}
\usepackage{pslatex}
\usepackage{ulem}
\usepackage{verbatim}
\usepackage{graphicx}
\usepackage{amssymb}
\usepackage{amsmath}
\usepackage{wrapfig}
\usepackage{float}
\usepackage{verbatim}
\usepackage{rotating}
\fontfamily{ptm}
\usepackage{hyperref}

\usepackage{pdfpages}

\usepackage{tocloft}
\usepackage{natbib}
\renewcommand{\icarus}{{\it Icarus}}

\newenvironment{itemize*}%
  {\begin{itemize}%
    \setlength{\itemsep}{3pt}%
    \setlength{\parskip}{0pt}}%
  {\end{itemize}}

\newenvironment{enumerate*}%
  {\begin{enumerate}%
    \setlength{\itemsep}{3pt}%
    \setlength{\parskip}{0pt}}%
  {\end{enumerate}}

\usepackage{color}
\usepackage{ulem}
\newcommand{\ACBc}[1]{\textcolor{black}{ #1}}

\begin{document}

\title{The {\it In Situ} Formation of Giant Planets at Short Orbital Periods}
\author{A.~C.~Boley\altaffilmark{1}
A.~P.~Granados Contreras\altaffilmark{1}, and
B.~Gladman\altaffilmark{1}}

\altaffiltext{1}{Department of Physics and Astronomy, The University of British Columbia, 6224 Agricultural Rd., Vancouver, B.C.~V6T 1Z4, Canada}

\begin{abstract}

We propose that two of the most surprising results so far among exoplanet discoveries are related: the existences of both hot Jupiters and the high frequency of multi-planet systems with periods $P\lesssim200$~days.
In this paradigm, the vast majority of stars rapidly form along with 
multiple close-in planets in the mass range of Mars to super-Earths/mini-Neptunes. 
Such systems of tightly-packed inner planets (STIPs) are metastable, with the time scale of the dynamical instability 
having a major influence on final planet types.
In most cases, the planets consolidate into a system of fewer, more massive planets,
but long after the circumstellar gas disk has dissipated.  This can yield
planets with masses above the traditional critical core of $\sim$10~$M_\oplus$, yielding short-period giants that lack abundant gas. 
A rich variety of physical states are also
possible given the range of collisional outcomes and formation time of the close-in planets.
However, when dynamical consolidation occurs before gas dispersal, a critical core can form that then grows via gas capture into a short-period
gas giant.
In this picture the majority of Hot and Warm Jupiters formed locally,
rather than migrating down from larger distances.

\end{abstract}

\maketitle

\section{INTRODUCTION}

The existence of Hot and Warm Jupiters (HJ and WJ, respectively)  has challenged planet formation theory since the discovery of these planet classes \citep{mayor_queloz_nature_1995,marcy_butler_aas_1995}.  
HJs are typically defined as planets with masses $\gtrsim 0.1 M_J$  on orbits with periods $P$ less than about 10 days  \citep[e.g.,][]{gaudi_etal_apj_2005,wright_etal_apj_2012}.
WJs are similar, but have orbits with longer periods out to the postulated water ice line ($\sim 1~\rm AU)$.
Together, HJs and WJs are members of a broader class of giant planets.
 We refer to any planet with a mass $>10 M_\oplus$ in the  HJ/WJ regions as a short-period giant (SPG).
We further use gSPG to distinguish any SPG that contains more than 50\% of its mass in hydrogen and helium.

About 10\% of FGK stars in the solar neighborhood harbor a giant planet at periods between 2 and 2000 days \citep{cumming_etal_pasp_2008,howard_science_2013}, with gSPGs comprising a large fraction of the sample. 
For example, approximately 1\% of FGK stars in the solar neighborhood harbor an HJ \citep{marcy_etal_ptps_2005,howard_etal_apjs_2012}, making the HJ frequency about 1/10 of the total giant frequency (for the given periods).
Under the classical paradigm of core nucleated instability \citep{pollack_etal_icarus_1996}, gas giant planet formation becomes favorable at nebular distances that are cold enough to allow water ice to condense. 
Because the water ice line is thought to occur at distances of at least 1 AU from solar-type stars, strict adherence to this classical paradigm requires that gas giant planets migrate over one or two orders of magnitude in semi-major axis to explain the SPG population.

Instead of migration,  one could question whether {\it in situ} gas giant formation is possible.  
This has been disfavored for  various reasons, including: (1) The amount of condensable solids in the nebula's  innermost region has been believed to be insufficient for reaching the critical core mass necessary for rapid gas capture  \citep{lin_etal_nature_1996}. However, this is based on popular models of the solar nebula, which may not be correct or may not reflect a general property of expoplanetary systems \citep[e.g.,][]{chiang_laughlin_mnras_2013}.
(2) The timescales required by core nucleated instability in the inner nebula are too long to explain the broad SPG population \citep{bodenheimer_etal_icarus_2000}.
(3) In the confirmed exoplanet catalogue, there had been a notable spike in the frequency of SPGs at periods of about three days (the ``3-day pileup'') followed by another rise in frequency near 1 AU \citep[e.g.,][]{cumming_etal_apj_1999,wright_etal_apj_2009,wright_exo}.  
This has been interpreted as evidence for migration, and is used as a test for different migration scenarios \citep[e.g.,][]{beauge_nesvorny_apj_2012}.


Here, we entertain the possibility that SPGs, including HJs, form {\it in situ}, motivated by advancements in exoplanet characterization and planet formation/evolution modeling.

 \begin{figure}
\includegraphics[width=0.45\textwidth]{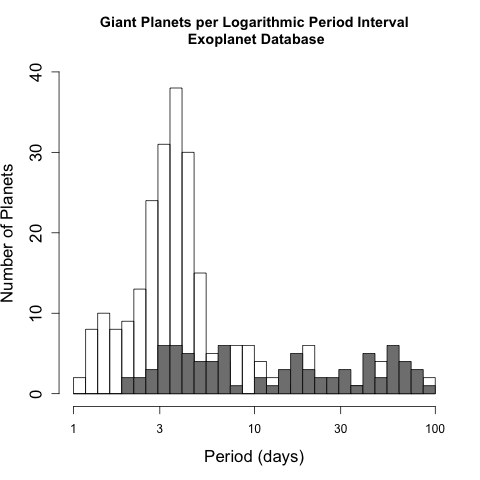}\includegraphics[width=0.45\textwidth]{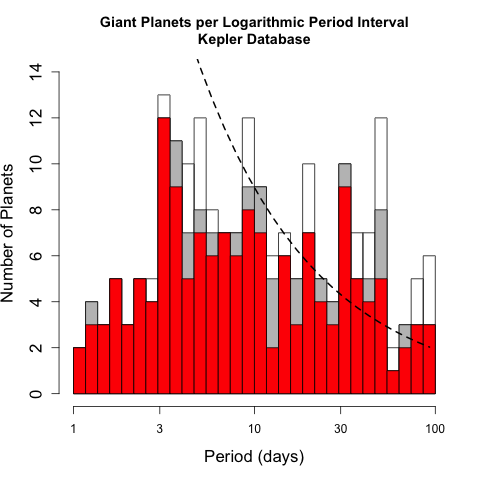}
\caption{Distribution functions for planets in the exoplanet.org database. Left: The white histogram shows all confirmed planets  with a measured or estimated size  $\rm 0.6 R_J <R <2 R_J$.  The gray histogram shows all planets with $M\sin i>0.1 M_J$ discovered by the radial velocity method.  Right:  Kepler planets and planet candidates for three radii cuts. 
Coloring denotes planet sizes between  $\rm 0.6 R_J <R <2 R_J$ (white), $\rm 0.7 R_J < R < 2 R_J$ (gray), and $\rm 0.8 R_J < R < 2 R_J$ (red), demonstrating that the trend is independent of the lower bound used to represent the gas giant threshold.eps
The 3-day pileup of HJs is only clearly present in the left panel white histogram, which is dominated by planets discovered by various transit surveys.  
In contrast, both the RV and  Kepler discoveries continue steadily after periods of 5 days (with a slow decrease), {\it despite increasing bias against detection at longer periods. }
To emphasize, the dependence of the single planet transit probability on orbital period $P$ is illustrated by the dashed curve ($\propto{ P}^{-2/3}$), assuming similar mass stars. 
The reason for this difference is not understood, as each survey has its own set of biases, but the above panels highlight that the 3-day pileup is not a general feature of  (g)SPG formation.
 \label{fig:perioddistro}}
\end{figure}

Figure \ref{fig:perioddistro} shows the period distributions  for planets and planet candidates with radii $R<2 R_J$ for three lower radii cuts, as well as $M\sin i>0.1 M_J$ for RV discoveries.  
These limits are chosen  to select probable gas giant planets. The left panel uses the exoplanets.org database \citep{wright_exo}, excluding Kepler planet candidates, while the right panel uses only the Kepler database\footnote{Data obtained via exoplanets.org, August 2015.}.
The 3-day pileup is present in the confirmed planets in the full exoplanet database, which is dominated by transit surveys. 
In contrast, the rapid drop in SPG frequency for $P>3$ days is not in the RV or Kepler data despite increasing bias against such detections.  
Moreover, using $\rm 0.6 R_J < R < 2 R_J$ for the Kepler sample,  about half of the candidates are HJs and the other half are WJs (for periods less than 100 days), with no obvious break in the distribution.
This highlights that the SPG formation mechanism does not, in general, produce a  3-day pileup.


We next consider Systems of Tightly-packed Inner Planets (STIPs),  found in abundance in the Kepler catalogue.
``Tightly-packed'' refers to the systems having multiple planets within the same period space as HJs/WJs. 
N-body simulations demonstrate that STIPs are prone to decay, likely due to secular dynamics \citep{lithwick_wu_pnas_2014,pu_wu_apj_2015,volk_gladman_apjl_2015}.
The  rate is logarithmic; roughly equal fraction of systems decay within equal logarithmic time intervals \citep{volk_gladman_apjl_2015}.
\ACBc{If STIPs form early in the presence of gas, then for the small fraction of systems that rapidly become unstable, consolidation could produce the critical cores necessary to initiate runaway gas capture, which is feasible at short periods \citep[e.g.,][]{bodenheimer_etal_icarus_2000}.
Moreover, massive super-Earths/mini-Neptunes are expected to accrete significant gas within disk lifetimes \citep{lee_etal_apj_2014}, regardless of their origin.
As such, high-density super-Earths and planets with super-critical masses that never became gas giants \citep{marcy_etal_2014}  are particularly challenging for the migration paradigm, as they should have accreted significant gas while migrating. }

Here we suggest that gas-poor SPGs arise from consolidation after gas is removed, while gSPGs, including HJs,  arise from the early consolidation of STIPs in the gaseous disk. 
We present n-body realizations of Kepler-11 as a case study to demonstrate the basic mechanism.  
We then place the results in a general picture that connects SPGs and low and high-density planets in STIPs to the conditions of the nebula at the time of STIP instability.

\section{Numerical Experiments}

We ran a series of n-body simulations to explore whether STIP decay could lead to SPG/gSPG formation.
The following framework  was used:
(1) STIPs typically form with high planet multiplicity. 
As such, planetary systems with lower multiplicity are the decay products of these initial systems. 
(2) The high multiplicity STIPs observed today are representative of initial formation configurations, albeit the longest-lived variants.  
(3) STIPs form quickly, producing many planets in the Mars to mini-Neptune range ($<10~M_\oplus$). 
While we expect a range of formation ages among STIPs, we assumed that all systems form well within 1 Myr of disk formation.  
We note that iron meteorite parent bodies in the Solar System formed within about 1.5 Myr of calcium-aluminum-rich inclusions \citep{schersten_etal_2006} and that HL Tau appears to be well into the throes of planet formation after  $\sim1$ Myr, even at very large orbital distances.  
This picture thus seems plausible, \ACBc{but not guaranteed, as such short formation timescales for super-Earths/mini-Neptunes require high surface densities \citep{dawson_etal_mnras_2015}.}
(4) The gaseous disk evolves rapidly, depleted on an exponential timescale of 2.5 Myr \citep[e.g.,][]{haisch_lada_lada,mamajek_2009}.
 (5) At early times, the natal disk has sufficient gas such that cores exceeding critical mass can grow to become gSPGs. 
 Finally, (6) the critical core mass occurs at a single mass ($10$ M$_\oplus$ here). 
In detail, the critical mass depends on the local disk conditions and duration that gas is available \citep{lee_chiang_apj_2015, piso_etal_apj_2015}, but the single mass allowed us to build a clean experiment.

 We ran 1000 realizations of the Kepler 11 system (K11), which is intended to represent a {\it plausible} initial STIP architecture, not the only configuration; \cite{volk_gladman_apjl_2015} studied a larger set.
We used  K11 because the planetary locations and masses are well constrained, except the mass of the outermost planet, K11g. 
We set the K11g mass to be 8 M$_\oplus$, which is comparable to other masses in the system and below both the derived upper mass limit \citep{lissauer_etal_nature_2011} and the assumed critical core mass. 
As such, the planet masses and semi-major axes (${\rm M_\oplus}$, AU)  are ${\rm K11b = (1.9,~0.09) }$ ${\rm K11c = (2.9,~0.11)}$, ${\rm K11d = (7.3,~0.15)}$, ${\rm K11e = (8,~0.19)}$, ${\rm K11f = (2,~0.25)}$, and ${\rm K11g = (8,~0.47)}$.
The pericenter argument, ascending node longitude, and mean anomaly are  uniformly distributed between 0 and 360 degrees. 
 The eccentricities are uniformly random between 0 and 0.05, and the inclinations are drawn from a Rayleigh distribution with dispersion $\sigma=1.8^\circ$ \citep{fabrycky_etal_apj_2014}.
 The inclination distribution is based on an analysis of the entire exoplanet population, while the eccentricity distribution is chosen for convenience in the absence of strong constraints on the low eccentricity regime.

 All simulations were run using Mercury6 \citep{mercury6}, modified to include a low-eccentricity correction for GR that captures the pericenter precession. 
 Each realization is run for 20 Myr, well beyond the lifetime for most gaseous disks. 
 We use the hybrid integrator, with a timestep of 0.1 days for the mixed-variable symplectic algorithm.  
 During close approaches, which occur when planets pass within one Hill radius of another, the Burlirsch-Stoer algorithm is used, with an accuracy parameter of $10^{-12}$.

 \section{Results}

\begin{figure}
\includegraphics[width=0.85\textwidth]{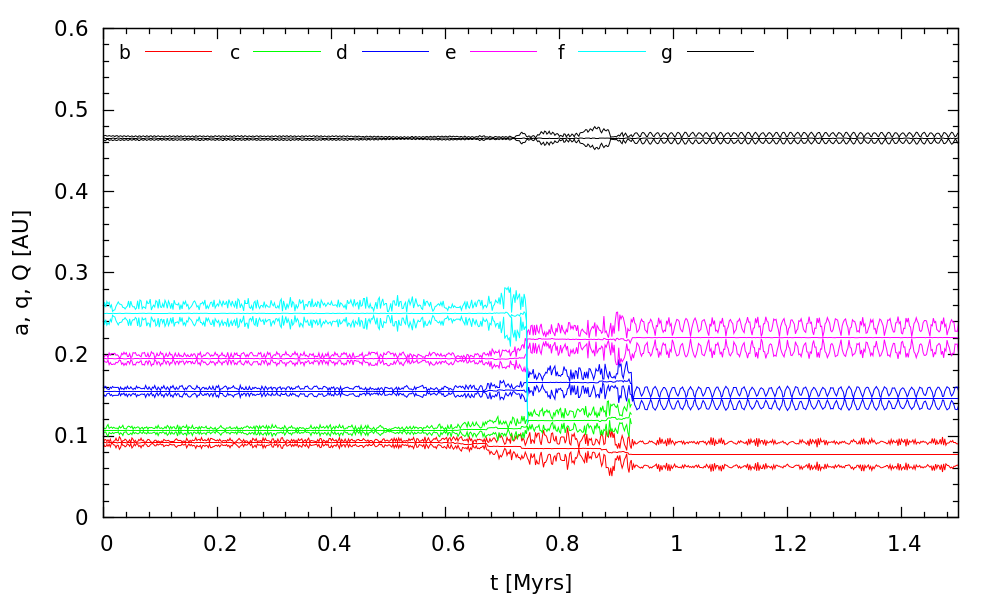}
\includegraphics[width=0.85\textwidth]{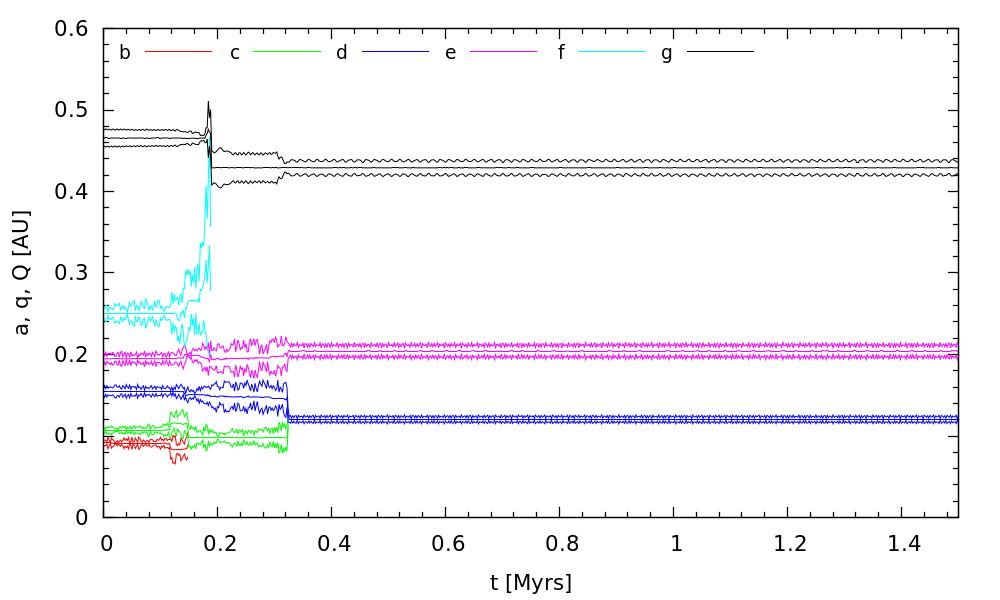}
\caption{ The semi-major axes $a$, pericentre $q$, and apocentre $Q$ as a function of time for two realizations of Kepler-11, exhibiting different dynamical evolutions. 
Top panel: the consolidation $({\rm [K11c+K11f] + K11d})$ of a STIP produces a four-planet system with one 12 $M_\oplus$ critical core (blue curve). 
Bottom panel: a three-planet system is produced $({\rm [K11b+K11c]+K11d})$ and $ ({\rm K11g + K11f})$ with two critical cores (black and blue curves). 
These cores are produced early enough in the disk's lifetime, assuming prompt STIP formation, that substantial gas could still be present.
 \label{fig:interactions}}
\end{figure}

\begin{figure}
\includegraphics[width=0.45\textwidth]{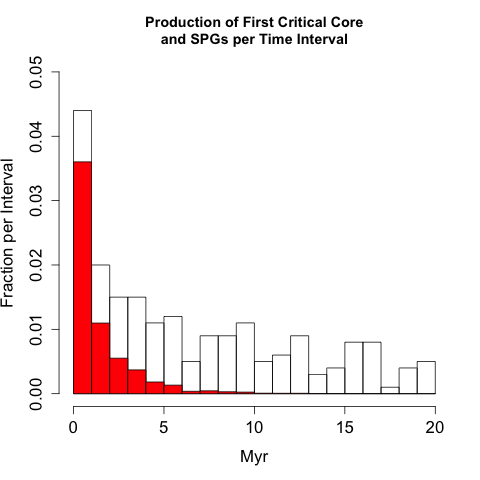}\includegraphics[width=0.45\textwidth]{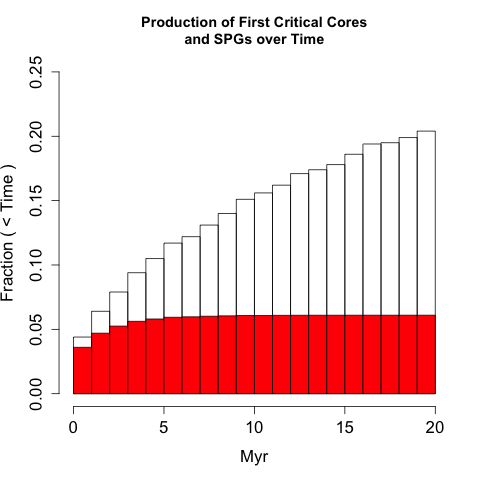}
\caption{The open histograms show the fraction of systems (cumulative and per time interval) that produce a critical core.  
The red histograms are similar, but with each bin weighted by $\exp(-\frac{t}{\rm 2.5~ Myr})$, where $t$ is the center of the given time interval.  
The exponential factor represents the observed decrease in disk accretion as a function of time, used here as a proxy for gas mass.
Thus, the red histograms reflect gaseous short-period giant (gSPG) formation.
 \label{fig:histograms_coll}}
\end{figure}

While the actual K11 is stable for Gyr timescales, of the 1000 realizations sampled here, 662 become unstable within 20 Myr.  
This large instability fraction is normal for STIPs \citep{volk_gladman_apjl_2015}.
Most mergers are between K11b and K11c, whose merged mass is below the $10\rm~M_\oplus$ threshold.
However, about 20\% of the realizations produce at least one critical core, with about 4.4\% of the collisions occurring within the first Myr. 
Two examples of consolidation outcomes are shown in Figure \ref{fig:interactions}.
In the first example, a four-planet system remains, with one critical core near the HJ period regime.  
In the second, two critical cores formed.
Figure \ref{fig:histograms_coll} highlights the fraction of systems that build at least one critical core, both per time interval and cumulative fraction. 

To estimate whether gSPG formation could proceed following the formation of critical cores by consolidation, we introduce a gas decay timescale of 2.5 Myr \citep{mamajek_2009}.
The total gas mass among protoplanetary disks varies, but even a minimum mass solar nebula contains significant gas mass at short periods, with disk accretion feeding additional gas to the region.
We set the start of the decay time ($t=0$) to be commensurate with STIP formation \citep[but see][]{dawson_etal_mnras_2015}.
In this model, $\exp\left(-\frac{t}{\rm 2.5~Myr}\right)$ is the fraction of disks at time $t$ that have enough gas to form a gSPG if a core appears. 
The red histograms in Figure  \ref{fig:histograms_coll}, representing gSPG formation,  are produced by weighting the open histograms in Figure  \ref{fig:histograms_coll}  by the disk fraction for the corresponding time interval center.
The cumulative fraction of systems that produce at least one gas giant over 20 Myr  is about 6\%, for our assumptions. 

After a critical core appears,  an additional $\sim 1$ Myr  may be required before runaway gas accretion can occur \citep{lee_etal_apj_2014}. 
Such a delay would decrease the above gSPG formation to about 4\%.
However, recent work \citep{pfalzner_etal_aa_2014} suggests that the disk exponential decay time is closer to 4 Myr.
Using this decay time with the 1 Myr delay brings the total gSPG formation back to about 6\%.  

\section{Discussion}

\begin{figure}
\includegraphics[width=0.85\textwidth]{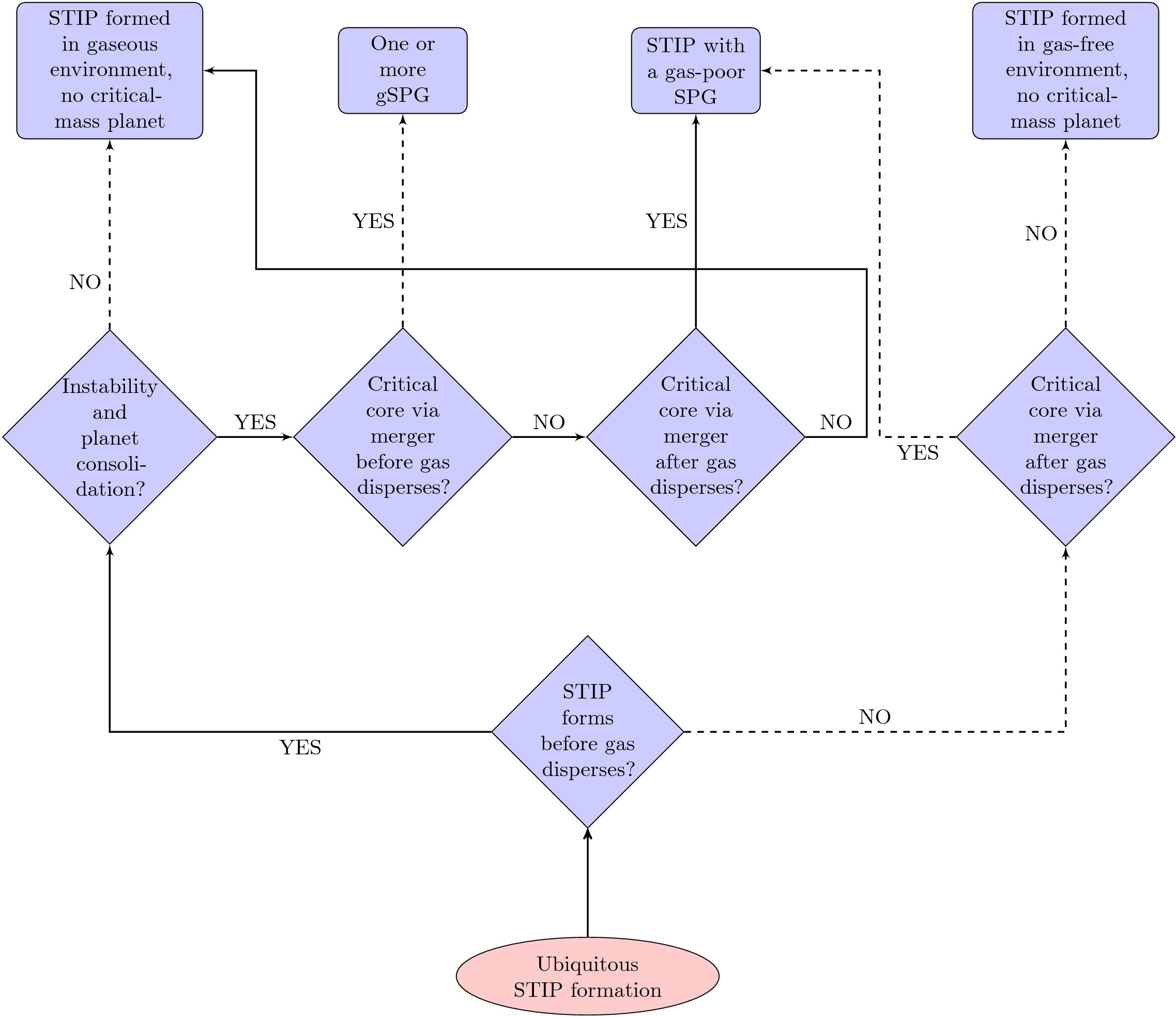}
\caption{Formation paradigm for the diversity of planets at short orbital periods.  
Dashed lines correspond to paths that are envisaged to be rarer than solid lines. 
In this picture, consolidation plays a major role in determining planet populations. 
gSPG formation is  possible if critical cores can form via system instability in the presence of gas.
In contrast, if consolidation takes place after the gas has dissipated, then an SPG could form without leading to gas giant planet formation. 
Bulk densities will be affected by initial STIP mass and whether the STIP formed in the presence of gas. 
Subsequent processes, such as irradiation and giant impacts, can increase the diversity of densities even further. 
 \label{fig:bigpicture}}
\end{figure}

The HJ frequency in the solar neighborhood is  $\sim1$\% \citep{wright_etal_apj_2012}, while it is about 0.5\% for the Kepler sample \citep{howard_etal_apjs_2012}.
The total Jupiter-type planet frequency around FGK stars in the solar neighborhood (for periods less than 2000 days) is about 10\% \citep{howard_science_2013}, based on RV surveys.
This includes giants at distances greater than 1 AU, where there is a rise in frequency per logarithmic period interval, possibly due to condensation of water ice (as classically viewed).
The calculations here yield a gSPG occurrence of approximately 6\%, comparable to the solar neighborhood gSPG frequency.
If STIP instability produces SPGs, then the time of instability relative to gas depletion may be a principal factor in leading to gSPG formation versus a system of super-Earths/mini-Neptunes.
This has been explored recently in the context of forming super-Earths with and without gas, where super-Earth/mini-Neptune formation in the presence of gas can lead to planets with low bulk densities (due to extended hydrogen atmospheres), while  
formation after gas dispersal can lead to higher-density planets \citep{dawson_etal_mnras_2015}, even those larger than the critical core mass.  
Subsequent evolution, such as collisional or stellar-induced atmosphere evaporation can further add to the density diversity. 
The overall picture is shown in Figure \ref{fig:bigpicture}.

Strictly, our K11 simulations do not produce a gas giant that would be classified as an HJ.
This is a direct consequence of using K11 as the analogue system, which has the most massive cores at larger semi-major axes (relative to the HJ threshold).
If the K11 planets were on slightly smaller orbits, then some of the gSPGs would potentially be HJs.
The purpose of these calculations is to demonstrate that local STIP instability could form critical cores, not specifically to match the period distribution, which requires knowing the unknown initial planet population.


Regardless of critical core production, local gSPG formation is only possible if sufficient gas is present. 
Consider the early solar nebula as an example, with a surface density  $\Sigma\propto r^{-1.5}$.
The combined gas and solid mass between $r_0$ and $r_1$ is 
$M(r_0,r_1) \approx 2.4 {\rm M_J} \left(\frac{\Sigma_0}{1700 {\rm g~cm^{-2}}}\right)\left(r_1^{0.5}-r_0^{0.5}\right)$, where $\Sigma_0$ is the surface density at 1 AU and distances are in AU.
The value $\Sigma_0\sim 1700 {\rm g~cm^{-2}}$ \citep{Hayashi_1981} is often used in the literature, but may be insufficient to form Jupiter without spatially concentrating solids above the nominal condensible solid abundance (e.g., Lodders 2003); $\Sigma_0$ closer to $\sim 4000 {\rm g~cm^{-2}}$ may be necessary for a {\it minimum} mass \citep{weidenschilling_mmsn, kuchner_apj_2004}.
Using this range for $\Sigma_0$, the total mass from 0.05 AU to 0.5 AU is between $\sim 1.2$ and $2.7~{\rm M_J}$ (about 2 and $4~{\rm M_{\oplus}}$ of condensible silicates and metals). 
If one uses the exoplanet population to estimate protoplanetary disk masses  \citep{chiang_laughlin_mnras_2013}, similar to that done for the solar nebula, then many systems had much higher available mass (in both solids and gas).
Local formation of gas giants could proceed provided that the gas reservoir can be delivered to the critical core. 
The detailed growth of the planet will depend on the mass accretion rate through the evolving disk and the interactions of the planet with that gas \cite[e.g.,][]{bodenheimer_etal_icarus_2000,benz_ppvi}. 
The inferred gas accretion rates among classical T Tauri stars is  $\dot{M}\sim 10^{-8}~ {\rm M_{\odot} yr^{-1}}$ \citep{gullbring_etal_apj_1998}, which is a mass flux of ${\rm 10~M_J~Myr^{-1}}$, suggesting significant gSPG growth is possible after initial formation.  

This mechanism does not intrinsically address the formation of large stellar spin-orbit misalignments \citep[e.g.,][]{albrecht_etal_apj_2012}. 
The cause of the misalignment is not understood, but may result from interactions between the disk and the star \citep{lai_etal_mnras_2011}, interactions between the disk and the system's birth environment \citep{bate_etal_mnras_2010,Batygin_nat_2012,fielding_etal_mnras_2015}, and/or dynamical interactions between the planets followed by tidal evolution \citep[e.g.,][]{nagasawa_ida_apj_2011,fabrycky_tremaine_apj_2007}.

Another potential problem is that while many gSPGs show evidence of having at least one companion, they are at much larger orbital distances \citep{knutson_etal_apj_2014}.  
As the collisional outcomes in Figure \ref{fig:interactions} show, super-Earth/mini-Neptune-type planets can be closely spaced to the critical core after its formation.  
However, as the core grows in mass to become a gSPG, subsequent dynamics may give rise to neighbor clearing.

The effects of gas-planet interactions are not included in these simulations.
Eccentricity damping in an isothermal disk \citep[e.g.,][]{tanaka_ward_apj_2004} suggests that the damping time (for small eccentricities) is less than 1 yr for a $\sim 10~\rm M_{\oplus}$ planet near 0.1 AU, depending on disk conditions.
At face value, this implies that our calculations overestimate the fraction of STIPs that can become unstable in the presence of gas. 
However, classic type I planet-disk interactions are known to be too efficient to explain planet populations without accounting for additional gas physics or processes \citep{baruteau_etal_ppvi,benz_ppvi}. 
If this also extends to eccentricity damping, then gas may be unable to completely prevent STIP instability.
This must be explored further.

Finally, the differences between the solar neighborhood and Kepler HJ frequencies, assuming both samples are representative, might be resolved by metallicity \citep{wright_etal_apj_2012}.  
A higher metallicity may result in initial STIPs that have higher masses or a higher planet multiplicity. 
It may also increase the fraction of stellar host disks that produce a STIP in the presence of significant gas, as high solid surface densities are needed for STIP formation before gas dissipation \citep{dawson_etal_mnras_2015}. 
We cautiously note that in the simulations presented here, there is a tendency for the inner planets to collide first \citep[noted by][who investigated additional STIPs]{volk_gladman_apjl_2015}.
If a metallicity increase enhances this tendency, then one expects a shift in the period distribution of critical cores toward shorter periods.  
Furthermore, in this picture, the increase in gas giant frequency at periods near 1 AU \citep{wright_etal_apj_2012} would be understood as the classical gas giant formation bias, where water ice contributes significantly in promoting critical core formation.  
Even in these outer regions, gas capture may be initiated through consolidation  of rocky-ice embryos.
No significant migration of massive planets is needed under this paradigm.
An additional consequence is that the presence of a gSPG does not preclude terrestrial planet formation near $\sim 1$ AU.



\ACBc{During the review of this manuscript, a complementary study by Batygin et al.~(2015) was posted to the arXiv.}

We thank Kat Volk, Scott Tremaine, and the anonymous referee for comments and discussions that improved this manuscript.
ACB was funded, in part, by the Canada Research Chairs program.
ACB and APGC acknowledge additional funding by The University of British Columbia and an NSERC Discovery grant.
BG was funded, in part, by an NSERC Discovery Grant. 
This research made use of the Exoplanet Orbit Database at exoplanets.org and was enabled in part by WestGrid and 
Compute Canada Calcul Canada.


\begin{thebibliography}{}
\expandafter\ifx\csname natexlab\endcsname\relax\def\natexlab#1{#1}\fi

\bibitem[{{Albrecht} {et~al.}(2012){Albrecht}, {Winn}, {Johnson}, {Howard},
  {Marcy}, {Butler}, {Arriagada}, {Crane}, {Shectman}, {Thompson}, {Hirano},
  {Bakos}, \& {Hartman}}]{albrecht_etal_apj_2012}
{Albrecht}, S., {Winn}, J.~N., {Johnson}, J.~A., {et~al.} 2012, \apj, 757, 18

\bibitem[{{Baruteau} {et~al.}(2014){Baruteau}, {Crida}, {Paardekooper},
  {Masset}, {Guilet}, {Bitsch}, {Nelson}, {Kley}, \&
  {Papaloizou}}]{baruteau_etal_ppvi}
{Baruteau}, C., {Crida}, A., {Paardekooper}, S.-J., {et~al.} 2014, Protostars
  and Planets VI, 667

\bibitem[{{Bate} {et~al.}(2010){Bate}, {Lodato}, \&
  {Pringle}}]{bate_etal_mnras_2010}
{Bate}, M.~R., {Lodato}, G., \& {Pringle}, J.~E. 2010, \mnras, 401, 1505

\bibitem[{{Batygin}(2012)}]{Batygin_nat_2012}
{Batygin}, K. 2012, \nat, 491, 418

\bibitem[{{Beaug{\'e}} \& {Nesvorn{\'y}}(2012)}]{beauge_nesvorny_apj_2012}
{Beaug{\'e}}, C., \& {Nesvorn{\'y}}, D. 2012, \apj, 751, 119

\bibitem[{{Benz} {et~al.}(2014){Benz}, {Ida}, {Alibert}, {Lin}, \&
  {Mordasini}}]{benz_ppvi}
{Benz}, W., {Ida}, S., {Alibert}, Y., {Lin}, D., \& {Mordasini}, C. 2014,
  Protostars and Planets VI, 691

\bibitem[{{Bodenheimer} {et~al.}(2000){Bodenheimer}, {Hubickyj}, \&
  {Lissauer}}]{bodenheimer_etal_icarus_2000}
{Bodenheimer}, P., {Hubickyj}, O., \& {Lissauer}, J.~J. 2000, \icarus, 143, 2

\bibitem[{{Chambers}(1999)}]{mercury6}
{Chambers}, J.~E. 1999, \mnras, 304, 793

\bibitem[{{Chiang} \& {Laughlin}(2013)}]{chiang_laughlin_mnras_2013}
{Chiang}, E., \& {Laughlin}, G. 2013, MNRAS, 431, 3444

\bibitem[{{Cumming} {et~al.}(2008){Cumming}, {Butler}, {Marcy}, {Vogt},
  {Wright}, \& {Fischer}}]{cumming_etal_pasp_2008}
{Cumming}, A., {Butler}, R.~P., {Marcy}, G.~W., {et~al.} 2008, PASP, 120, 531

\bibitem[{{Cumming} {et~al.}(1999){Cumming}, {Marcy}, \&
  {Butler}}]{cumming_etal_apj_1999}
{Cumming}, A., {Marcy}, G.~W., \& {Butler}, R.~P. 1999, \apj, 526, 890

\bibitem[{{Dawson} {et~al.}(2015){Dawson}, {Chiang}, \&
  {Lee}}]{dawson_etal_mnras_2015}
{Dawson}, R.~I., {Chiang}, E., \& {Lee}, E.~J. 2015, \mnras, 453, 1471

\bibitem[{{Fabrycky} \& {Tremaine}(2007)}]{fabrycky_tremaine_apj_2007}
{Fabrycky}, D., \& {Tremaine}, S. 2007, \apj, 669, 1298

\bibitem[{{Fabrycky} {et~al.}(2014){Fabrycky}, {Lissauer}, {Ragozzine}, {Rowe},
  {Steffen}, {Agol}, {Barclay}, {Batalha}, {Borucki}, {Ciardi}, {Ford},
  {Gautier}, {Geary}, {Holman}, {Jenkins}, {Li}, {Morehead}, {Morris},
  {Shporer}, {Smith}, {Still}, \& {Van Cleve}}]{fabrycky_etal_apj_2014}
{Fabrycky}, D.~C., {Lissauer}, J.~J., {Ragozzine}, D., {et~al.} 2014, \apj,
  790, 146

\bibitem[{{Fielding} {et~al.}(2015){Fielding}, {McKee}, {Socrates},
  {Cunningham}, \& {Klein}}]{fielding_etal_mnras_2015}
{Fielding}, D.~B., {McKee}, C.~F., {Socrates}, A., {Cunningham}, A.~J., \&
  {Klein}, R.~I. 2015, \mnras, 450, 3306

\bibitem[{{Gaudi} {et~al.}(2005){Gaudi}, {Seager}, \&
  {Mallen-Ornelas}}]{gaudi_etal_apj_2005}
{Gaudi}, B.~S., {Seager}, S., \& {Mallen-Ornelas}, G. 2005, \apj, 623, 472

\bibitem[{{Gullbring} {et~al.}(1998){Gullbring}, {Hartmann}, {Brice{\~n}o}, \&
  {Calvet}}]{gullbring_etal_apj_1998}
{Gullbring}, E., {Hartmann}, L., {Brice{\~n}o}, C., \& {Calvet}, N. 1998, \apj,
  492, 323

\bibitem[{{Haisch} {et~al.}(2001){Haisch}, {Lada}, \&
  {Lada}}]{haisch_lada_lada}
{Haisch}, Jr., K.~E., {Lada}, E.~A., \& {Lada}, C.~J. 2001, \apjl, 553, L153

\bibitem[{{Hayashi}(1981)}]{Hayashi_1981}
{Hayashi}, C. 1981, Progress of Theoretical Physics Supplement, 70, 35

\bibitem[{{Howard}(2013)}]{howard_science_2013}
{Howard}, A.~W. 2013, Science, 340, 572

\bibitem[{{Howard} {et~al.}(2012){Howard}, {Marcy}, {Bryson}, {Jenkins},
  {Rowe}, {Batalha}, {Borucki}, {Koch}, {Dunham}, {Gautier}, {Van Cleve},
  {Cochran}, {Latham}, {Lissauer}, {Torres}, {Brown}, {Gilliland}, {Buchhave},
  {Caldwell}, {Christensen-Dalsgaard}, {Ciardi}, {Fressin}, {Haas}, {Howell},
  {Kjeldsen}, {Seager}, {Rogers}, {Sasselov}, {Steffen}, {Basri},
  {Charbonneau}, {Christiansen}, {Clarke}, {Dupree}, {Fabrycky}, {Fischer},
  {Ford}, {Fortney}, {Tarter}, {Girouard}, {Holman}, {Johnson}, {Klaus},
  {Machalek}, {Moorhead}, {Morehead}, {Ragozzine}, {Tenenbaum}, {Twicken},
  {Quinn}, {Isaacson}, {Shporer}, {Lucas}, {Walkowicz}, {Welsh}, {Boss},
  {Devore}, {Gould}, {Smith}, {Morris}, {Prsa}, {Morton}, {Still}, {Thompson},
  {Mullally}, {Endl}, \& {MacQueen}}]{howard_etal_apjs_2012}
{Howard}, A.~W., {Marcy}, G.~W., {Bryson}, S.~T., {et~al.} 2012, \apjs, 201, 15

\bibitem[{{Knutson} {et~al.}(2014){Knutson}, {Fulton}, {Montet}, {Kao}, {Ngo},
  {Howard}, {Crepp}, {Hinkley}, {Bakos}, {Batygin}, {Johnson}, {Morton}, \&
  {Muirhead}}]{knutson_etal_apj_2014}
{Knutson}, H.~A., {Fulton}, B.~J., {Montet}, B.~T., {et~al.} 2014, \apj, 785,
  126

\bibitem[{{Kuchner}(2004)}]{kuchner_apj_2004}
{Kuchner}, M.~J. 2004, \apj, 612, 1147

\bibitem[{{Lai} {et~al.}(2011){Lai}, {Foucart}, \& {Lin}}]{lai_etal_mnras_2011}
{Lai}, D., {Foucart}, F., \& {Lin}, D.~N.~C. 2011, \mnras, 412, 2790

\bibitem[{{Lee} \& {Chiang}(2015)}]{lee_chiang_apj_2015}
{Lee}, E.~J., \& {Chiang}, E. 2015, \apj, 811, 41

\bibitem[{{Lee} {et~al.}(2014){Lee}, {Chiang}, \& {Ormel}}]{lee_etal_apj_2014}
{Lee}, E.~J., {Chiang}, E., \& {Ormel}, C.~W. 2014, \apj, 797, 95

\bibitem[{{Lin} {et~al.}(1996){Lin}, {Bodenheimer}, \&
  {Richardson}}]{lin_etal_nature_1996}
{Lin}, D.~N.~C., {Bodenheimer}, P., \& {Richardson}, D.~C. 1996, \nat, 380, 606

\bibitem[{{Lissauer} {et~al.}(2011){Lissauer}, {Fabrycky}, {Ford}, {Borucki},
  {Fressin}, {Marcy}, {Orosz}, {Rowe}, {Torres}, {Welsh}, {Batalha}, {Bryson},
  {Buchhave}, {Caldwell}, {Carter}, {Charbonneau}, {Christiansen}, {Cochran},
  {Desert}, {Dunham}, {Fanelli}, {Fortney}, {Gautier}, {Geary}, {Gilliland},
  {Haas}, {Hall}, {Holman}, {Koch}, {Latham}, {Lopez}, {McCauliff}, {Miller},
  {Morehead}, {Quintana}, {Ragozzine}, {Sasselov}, {Short}, \&
  {Steffen}}]{lissauer_etal_nature_2011}
{Lissauer}, J.~J., {Fabrycky}, D.~C., {Ford}, E.~B., {et~al.} 2011, Nature,
  470, 53

\bibitem[{{Lithwick} \& {Wu}(2014)}]{lithwick_wu_pnas_2014}
{Lithwick}, Y., \& {Wu}, Y. 2014, Proceedings of the National Academy of
  Science, 111, 12610

\bibitem[{{Mamajek}(2009)}]{mamajek_2009}
{Mamajek}, E.~E. 2009, in American Institute of Physics Conference Series, Vol.
  1158, American Institute of Physics Conference Series, ed. T.~{Usuda},
  M.~{Tamura}, \& M.~{Ishii}, 3--10

\bibitem[{{Marcy} {et~al.}(2005){Marcy}, {Butler}, {Fischer}, {Vogt}, {Wright},
  {Tinney}, \& {Jones}}]{marcy_etal_ptps_2005}
{Marcy}, G., {Butler}, R.~P., {Fischer}, D., {et~al.} 2005, Progress of
  Theoretical Physics Supplement, 158, 24

\bibitem[{{Marcy} \& {Butler}(1995)}]{marcy_butler_aas_1995}
{Marcy}, G.~W., \& {Butler}, R.~P. 1995, in Bulletin of the American
  Astronomical Society, Vol.~27, American Astronomical Society Meeting
  Abstracts, 1379

\bibitem[{{Marcy} {et~al.}(2014){Marcy}, {Isaacson}, {Howard}, {Rowe},
  {Jenkins}, {Bryson}, {Latham}, {Howell}, {Gautier}, {Batalha}, {Rogers},
  {Ciardi}, {Fischer}, {Gilliland}, {Kjeldsen}, {Christensen-Dalsgaard},
  {Huber}, {Chaplin}, {Basu}, {Buchhave}, {Quinn}, {Borucki}, {Koch}, {Hunter},
  {Caldwell}, {Van Cleve}, {Kolbl}, {Weiss}, {Petigura}, {Seager}, {Morton},
  {Johnson}, {Ballard}, {Burke}, {Cochran}, {Endl}, {MacQueen}, {Everett},
  {Lissauer}, {Ford}, {Torres}, {Fressin}, {Brown}, {Steffen}, {Charbonneau},
  {Basri}, {Sasselov}, {Winn}, {Sanchis-Ojeda}, {Christiansen}, {Adams},
  {Henze}, {Dupree}, {Fabrycky}, {Fortney}, {Tarter}, {Holman}, {Tenenbaum},
  {Shporer}, {Lucas}, {Welsh}, {Orosz}, {Bedding}, {Campante}, {Davies},
  {Elsworth}, {Handberg}, {Hekker}, {Karoff}, {Kawaler}, {Lund}, {Lundkvist},
  {Metcalfe}, {Miglio}, {Silva Aguirre}, {Stello}, {White}, {Boss}, {Devore},
  {Gould}, {Prsa}, {Agol}, {Barclay}, {Coughlin}, {Brugamyer}, {Mullally},
  {Quintana}, {Still}, {Thompson}, {Morrison}, {Twicken}, {D{\'e}sert},
  {Carter}, {Crepp}, {H{\'e}brard}, {Santerne}, {Moutou}, {Sobeck}, {Hudgins},
  {Haas}, {Robertson}, {Lillo-Box}, \& {Barrado}}]{marcy_etal_2014}
{Marcy}, G.~W., {Isaacson}, H., {Howard}, A.~W., {et~al.} 2014, \apjs, 210, 20

\bibitem[{{Mayor} \& {Queloz}(1995)}]{mayor_queloz_nature_1995}
{Mayor}, M., \& {Queloz}, D. 1995, \nat, 378, 355

\bibitem[{{Nagasawa} \& {Ida}(2011)}]{nagasawa_ida_apj_2011}
{Nagasawa}, M., \& {Ida}, S. 2011, \apj, 742, 72

\bibitem[{{Pfalzner} {et~al.}(2014){Pfalzner}, {Steinhausen}, \&
  {Menten}}]{pfalzner_etal_aa_2014}
{Pfalzner}, S., {Steinhausen}, M., \& {Menten}, K. 2014, \apjl, 793, L34

\bibitem[{{Piso} {et~al.}(2015){Piso}, {Youdin}, \&
  {Murray-Clay}}]{piso_etal_apj_2015}
{Piso}, A.-M.~A., {Youdin}, A.~N., \& {Murray-Clay}, R.~A. 2015, \apj, 800, 82

\bibitem[{{Pollack} {et~al.}(1996){Pollack}, {Hubickyj}, {Bodenheimer},
  {Lissauer}, {Podolak}, \& {Greenzweig}}]{pollack_etal_icarus_1996}
{Pollack}, J.~B., {Hubickyj}, O., {Bodenheimer}, P., {et~al.} 1996, \icarus,
  124, 62

\bibitem[{{Pu} \& {Wu}(2015)}]{pu_wu_apj_2015}
{Pu}, B., \& {Wu}, Y. 2015, \apj, 807, 44

\bibitem[{{Scherst{\'e}n} {et~al.}(2006){Scherst{\'e}n}, {Elliott},
  {Hawkesworth}, {Russell}, \& {Masarik}}]{schersten_etal_2006}
{Scherst{\'e}n}, A., {Elliott}, T., {Hawkesworth}, C., {Russell}, S., \&
  {Masarik}, J. 2006, Earth and Planetary Science Letters, 241, 530

\bibitem[{{Tanaka} \& {Ward}(2004)}]{tanaka_ward_apj_2004}
{Tanaka}, H., \& {Ward}, W.~R. 2004, \apj, 602, 388

\bibitem[{{Volk} \& {Gladman}(2015)}]{volk_gladman_apjl_2015}
{Volk}, K., \& {Gladman}, B. 2015, \apjl, 806, L26

\bibitem[{{Weidenschilling}(1977)}]{weidenschilling_mmsn}
{Weidenschilling}, S.~J. 1977, \apss, 51, 153

\bibitem[{{Wright} {et~al.}(2012){Wright}, {Marcy}, {Howard}, {Johnson},
  {Morton}, \& {Fischer}}]{wright_etal_apj_2012}
{Wright}, J.~T., {Marcy}, G.~W., {Howard}, A.~W., {et~al.} 2012, \apj, 753, 160

\bibitem[{{Wright} {et~al.}(2009){Wright}, {Upadhyay}, {Marcy}, {Fischer},
  {Ford}, \& {Johnson}}]{wright_etal_apj_2009}
{Wright}, J.~T., {Upadhyay}, S., {Marcy}, G.~W., {et~al.} 2009, \apj, 693, 1084

\bibitem[{{Wright} {et~al.}(2011){Wright}, {Fakhouri}, {Marcy}, {Han}, {Feng},
  {Johnson}, {Howard}, {Fischer}, {Valenti}, {Anderson}, \&
  {Piskunov}}]{wright_exo}
{Wright}, J.~T., {Fakhouri}, O., {Marcy}, G.~W., {et~al.} 2011, PASP, 123, 412

\end{thebibliography}

\end{document}